\documentclass[5p,times]{elsarticle}
\usepackage{graphicx}
\usepackage{amsmath}
\usepackage[english]{babel}
\usepackage{amsfonts,amssymb}
\usepackage[mathscr]{eucal}

\addto\captionsenglish{\renewcommand{\appendixname}{Appendix }}

\makeatletter
\def\ps@pprintTitle{%
  \let\@oddhead\@empty
  \let\@evenhead\@empty
  \def\@oddfoot{\reset@font\hfil\thepage\hfil}
  \let\@evenfoot\@oddfoot
}
\makeatother

\usepackage[colorlinks=true]{hyperref}

\begin{document}

\title{Magnetic anisotropy of polycrystalline high-temperature\\
ferromagnetic Mn$_x$Si$_{1-x}$ ($x\approx0.5$) alloy films}

\author[ipp]{A.B.~Drovosekov\corref{cor}}
\ead{drovosekov@kapitza.ras.ru}
\author[ipp]{N.M.~Kreines}
\author[ipp]{A.O.~Savitsky}
\author[nrc,ipt]{S.V.~Kapelnitsky}
\author[nrc,ire]{V.V.~Rylkov}
\author[nrc,gpi]{V.V.~Tugushev}
\author[nrc]{G.V.~Prutskov}
\author[ilit]{O.A.~Novodvorskii}
\author[ilit]{A.V.~Shorokhova}
\author[hzd]{Y.~Wang}
\author[hzd]{S.~Zhou}

\cortext[cor]{Corresponding author}

\address[ipp]{P.L.Kapitza Institute for Physical Problems RAS, Kosygina St. 2, 
119334 Moscow, Russia}

\address[nrc]{National Research Centre "Kurchatov Institute", Kurchatov Sq. 1, 
123182 Moscow, Russia}

\address[ipt]{Institute of Physics and Technology RAS, Nakhimovsky av. 36, build 
1, 117218 Moscow, Russia}

\address[ire]{Kotel'nikov Institute of Radio Engineering and Electronics RAS, 
141190 Fryazino, Moscow region, Russia}

\address[gpi]{Prokhorov General Physics Institute RAS, Vavilov St. 38, 119991 
Moscow, Russia}

\address[ilit]{Institute on Laser and Information Technologies RAS, 
Svyatoozerskaya St. 1, 140700 Shatura,  Moscow region, Russia}

\address[hzd] {Helmholtz-Zentrum Dresden-Rossendorf, Institute of Ion Beam 
Physics and Materials Research, Bautzner Landstrasse 400, 01328 Dresden, 
Germany}

\begin{abstract}
A set of thin film Mn$_x$Si$_{1-x}$ alloy samples with different manganese 
concentration $x\approx0.44-0.63$ grown by the pulsed laser deposition (PLD) 
method onto the Al$_2$O$_3$~(0001) substrate was investigated in the temperature 
range $4-300$~K using ferromagnetic resonance (FMR) measurements in the wide 
range of frequencies ($f = 7 - 60$~GHz) and magnetic fields ($H = 0 -30$~kOe). 
For samples with $x\approx0.52-0.55$, FMR data show clear evidence of 
ferromagnetism with high Curie temperatures $T_\text{C} \sim 300$~K. 
These samples demonstrate complex and unusual character of magnetic anisotropy 
described in the frame of phenomenological model as a combination of the essential 
second order easy plane anisotropy contribution and the additional forth order 
uniaxial anisotropy contribution with easy direction normal to the film plane. 
We explain the obtained results by a polycrystalline (mosaic) structure of the 
films caused by the film-substrate lattice mismatch. The existence 
of extra strains at the crystallite boundaries leads to an essential inhomogeneous 
magnetic anisotropy in the film plane.
\end{abstract}


\maketitle

\section{Introduction}

Development of Si based magnetic semiconductor materials for spintronic 
applications attracts a lot of attention, since these materials can be 
easily incorporated into the existing microelectronic technology \cite{1}. In 
particular, Si-Mn alloys demonstrating unusual magnetic and transport properties 
[2--8] have especial interest to engineer non-conventional 
integrated-circuit elements.

However, there exist significant technological and fundamental obstacles to 
adapt Si-Mn based elements to the needs of spintronic. At relatively low Mn 
content in Mn$_x$Si$_{1-x}$ alloys ($x = 0.05 - 0.1$), the ferromagnetism (FM) 
at above room temperature has been revealed. But these alloys turn out to be 
strongly inhomogeneous materials due to their phase segregation, leading to 
formation of isolated magnetic MnSi$_{1.7}$ precipitate nanoparticles with the 
Mn content $x \approx 0.35$ in Si matrix \cite{2}. In such alloys, anomalous 
Hall effect testifying the spin polarization of carriers is absent, that 
makes impossible to use these materials in spintronic applications. At the same 
time, the fabrication of well-reproducible homogeneous magnetic alloys with high 
Mn content $x \approx 0.35$ is difficult because of the variety of stable phases 
of high MnSi$_y$ silicides (not less than five) with the close content of 
components ($y = 1.72 - 1.75$) \cite{6,7}.

In contrast, nonstoichiometric Mn$_x$Si$_{1-x}$ alloys with high Mn content ($x 
\approx 0.5$, i.e. close to stoichiometric MnSi) are not inclined to a phase 
segregation and formation of isolated magnetic precipitates, so they seem more 
promising for spintronic applications than dilute Mn$_x$Si$_{1-x}$ alloys. 
Recently we have found that in thin films of such concentrated alloys, the Curie 
temperature $T_\text{C}$ increases by more than an order of magnitude as 
compared with bulk MnSi ($T_\text{C} \approx 30$~K) \cite{6}. Comparative 
studies of anomalous Hall effect and transverse Kerr effect showed that the 
ferromagnetic transition in Mn$_x$Si$_{1-x}$ ($x \approx 0.52 - 0.55$) alloys 
occurring at $T \sim 300$~K, has a global nature and is not associated with the 
phase segregation \cite{7}. Besides high $T_\text{C}$ values, the films 
investigated in \cite{6,7} show large values of saturation magnetization 
reaching $\approx 400$~emu/cm$^3$ at low temperatures. The observed 
magnetization value corresponds to $\approx 1.1 ~\mu_B$/Mn, that significantly 
exceeds the value $0.4 ~\mu_B$/Mn typical for bulk MnSi crystal \cite{9}.

High temperature FM in Mn$_x$Si$_{1-x}$ ($x \approx 0.5$) alloys has been 
qualitatively interpreted \cite{6,7} in frame of the early proposed model for 
dilute Mn$_x$Si$_{1-x}$ alloys \cite{3}, i.e. in terms of complex defects  with 
local magnetic moments embedded into the matrix of itinerant FM. However, many 
details of FM order in Mn$_x$Si$_{1-x}$ ($x \approx 0.5$) alloys are still not 
completely clear due to insufficient experimental studies. In particular, there 
are no data on their magnetic anisotropy features. In the present work, thin 
Mn$_x$Si$_{1-x}$ ($x \approx 0.52 - 0.55$) films are investigated by the 
ferromagnetic resonance (FMR) method which is powerful for providing valuable 
information about magnetic anisotropy peculiarities of thin film magnetic 
materials, in particular, like dilute magnetic semiconductors (see \cite{10} and 
references therein). Our studies are largely focused on the 
position and shape of FMR signal, while the analysis of the refinements of the 
line width data providing additional information on the magnetic inhomogeneity 
and relaxation rate of magnetization is not presented in this work. Besides, to 
study the details of crystalline and magnetic microstructures of our films we 
perform in this work the atomic force microscopy (AFM) and magnetic force 
microscopy (MFM) investigations. We hope that AFM and MFM methods allow for 
additional understanding of the FMR results, since these methods are able to 
reveal the "local" effects of the crystal and magnetic texture of the film on 
the origin of magnetic anisotropy established in FMR measurements.

\section{Samples and experimental details}

We studied six samples with manganese content in the range $x \approx 0.44 - 
0.63$. The 70~nm thick film samples were produced by the pulse laser deposition 
(PLD) method on Al$_2$O$_3$~(0001) substrates at $340\,^\circ$C.
The composition of the films was testified by X-ray photoelectronic spectroscopy 
(for details see \cite{6}).

The structural properties of the samples were studied by X-ray diffraction (XRD) 
analysis using a Rigaku SmartLab diffractometer without monochromators and a diaphragm 
before the detector. In this case, intensity of the direct beam was as high as 
$1.5\cdot10^9$~pulses/s. Additionally, we performed room temperature AFM and MFM 
investigations in the sample Mn$_x$Si$_{1-x}$ ($x \approx 0.52$) having the most 
pronounced high-$T_\text{C}$ FM, using microscope SmartSPM (AIST-NT).

FMR spectra of the obtained samples were studied at temperatures $T = 4 - 300$~K 
in the wide range of frequencies ($f = 7 - 60$~GHz) and magnetic fields ($H = 0 
- 30$~kOe). To detect the resonance absorption signal, magnetic field 
dependencies of the microwave power transmitted through the cavity resonator with 
the sample inside were measured at constant frequency. Measurements were carried 
out for different orientations of the magnetic field with respect to the film 
plane.

\section{Experimental results and discussion}

\subsection{FMR measurements}

\begin{figure}
\centering
\includegraphics[width=0.9\columnwidth]{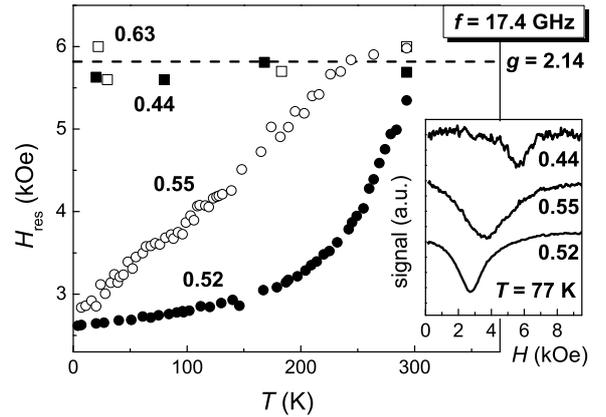}
\caption{Temperature dependence of the resonance field $H_\text{res}(T)$ at the 
frequency 17.4~GHz for samples with different manganese concentration. The 
dashed line corresponds to the calculated position of paramagnetic resonance for 
g-factor $g = 2.14$. The inset demonstrates examples of the experimental 
resonance spectra at $T = 77$~K. The field is applied in the film plane.}
\label{fig1}
\end{figure}

The dependence of the resonance field on temperature in case of the field 
applied in the film plane is shown in Fig.\ref{fig1} for several samples. Films with Mn 
concentration $x \approx 0.44$ and $x \approx 0.63$ demonstrate the resonance 
absorption peak in the field corresponding to a paramagnetic resonance 
situation, i.e. $f = \gamma H$, with the gyromagnetic ratio $\gamma \approx 
3$~GHz/kOe corresponding to the g-factor value $g = 2.14$, which  is in 
agreement with the value reported in Ref.~\cite{11} for the bulk MnSi single 
crystal. The paramagnetism of the samples with $x \approx 0.44$ and $x \approx 
0.63$ is observed in the temperature range $20 - 300$~K, that is in accordance 
with the results of Ref.~\cite{6}.

At low temperatures in the range of concentrations $x \approx 0.52 - 0.55$, the 
films show significant shift of absorption peak into the region of smaller 
fields with respect to paramagnetic samples (Fig.\ref{fig1}). As the temperature 
increases, the resonance field also increases and at $T \sim 300$~K reaches the 
value corresponding to the paramagnetic resonance situation. The observed 
absorption lines have the Lorentz-like shape, and the resonance field anisotropy 
in the film plane is absent.

The shift of the FMR absorption line revealed in our measurements (shown in 
Fig.\ref{fig1}) clearly indicates the presence of FM moment in samples with $x \approx 
0.52 - 0.55$ at low temperatures, in agreement with Ref.~\cite{6}. When the 
external magnetic field $H$ is applied in the plane of a thin FM film with 
magnetization $M$, the FMR frequency $f_\parallel$ may be described by the 
phenomenological formula (see \ref{appA}):
\begin{equation}
f_\parallel=\gamma\sqrt{H(H+K_\parallel M)}.
\label{fmr1}
\end{equation}
Here $K_\parallel$ is the effective easy plane magnetic anisotropy coefficient; 
$K_\parallel M$ is the effective field of easy plane magnetic anisotropy. This 
field describes the effect of two factors on the FMR spectra: 1)~the shape 
magnetic anisotropy depending of the form of the sample and 2)~the crystal 
structure driven magnetic anisotropy which is due to relativistic interactions 
between electrons and ions in the sample material. Following a simplest model 
used in Ref.~\cite{12} it is easy to obtain:
\begin{equation}
K_\parallel M=4\pi M+\frac{2K_1}{M},
\label{fmr1a}
\end{equation}
where $K_1$ is the phenomenological constant of the second-order easy plane 
crystal structure driven magnetic anisotropy. In the absence of $K_1$, Eq.(\ref{fmr1}) 
transforms into the well known Kittel formula for the applied field $H$ lying in 
the film plane \cite{13}.

According to Eq.(\ref{fmr1}), a reduction of FM moment of the film with increasing 
temperature leads to a decrease of the FMR line shift with respect to the 
paramagnetic resonance situation. Thus FMR data (Fig.\ref{fig1}) confirm the FM order up 
to $T \sim 300$~K in samples with Mn concentration $x \approx 0.52 - 0.55$.

\begin{figure}
\centering
\includegraphics[width=0.85\columnwidth]{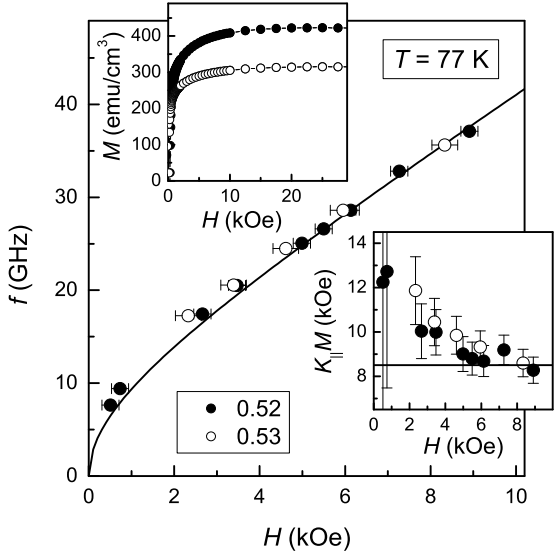}
\caption{Dependence of the resonance frequency on the magnetic field applied in 
the film plane for samples with $x \approx 0.52$ and $x \approx 0.53$ at $T = 
77$~K. Points are experimental data; the line is the theoretical curve according 
to Eq.(\protect\ref{fmr1}). The insets represent static magnetization curves (the upper 
inset) and the value of the parameter $K_\parallel M$ in Eq.(\protect\ref{fmr1}) as a 
function of magnetic field (the bottom inset).}
\label{fig2}
\end{figure}

Dependencies of the FMR frequency on the magnetic field applied in the film plane 
for samples with Mn concentrations 0.52 and 0.53 at $T=77$~K are shown in 
Fig.\ref{fig2}. At sufficiently high fields $H > 5$~kOe (i.e. in the high-frequency 
region $f > 25$~GHz), the $f(H)$ dependence can be well approximated by Eq.(\ref{fmr1}) 
with the field-independent value $K_\parallel M \approx 8.7$~kOe (at $T=77$~K 
both samples have about the same $K_\parallel M$ values). However, some 
deviations of the experimental points from the theoretical curve are observed at 
smaller fields. In the bottom inset of Fig.\ref{fig2}, the experimentally obtained field 
dependence of $K_\parallel M$ parameter is shown. One can see that the value of 
$K_\parallel M$ parameter significantly depends on the applied field at small 
fields $H < 5$~kOe, while it comes nigh unto a saturation at higher fields $H > 
5$~kOe.

The observed deviation of the $K_\parallel M$ parameter at low fields from the 
constant value at high fields is probably due to a random distribution of 
local magnetic anisotropy axes in the sample plane resulting in inhomogeneity of 
magnetization at low applied fields. Indeed, static 
magnetization curves (upper inset in Fig.\ref{fig2}) achieve a saturation in relatively 
high fields $H \sim 5$~kOe. One can notice, however, that the $K_\parallel M$ 
parameter estimated from the FMR data increases as the magnetic field decreases 
below 5~kOe, while the static magnetization diminishes. This contradiction 
clearly shows inapplicability of Eq.(\ref{fmr1}) in the region of small fields 
 and indicates inhomogeneity of local magnetization and anisotropy. 
Below in this work, only FMR data obtained at high frequencies $f > 25$~GHz are 
taken into account to estimate the magnetic anisotropy parameters of the system. 
Such frequencies provide sufficiently high resonance fields $H > 5$~kOe, where 
the FMR data can be well approximated in frame of Kittel's formalism (Eq.(\ref{fmr1})).

\begin{figure*}
\centering
\includegraphics[width=.95\textwidth]{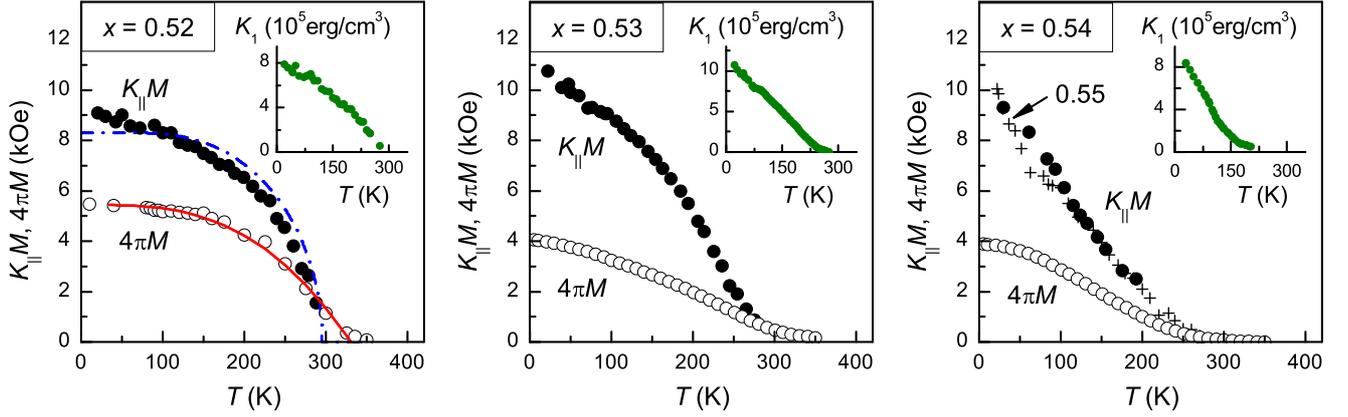}
\caption{Temperature dependence of $K_\parallel M$ parameter 
(solid circles) obtained from FMR, and demagnetizing field $4\pi M$ 
(open circles) obtained from static magnetization data for samples 
with concentration $x \approx 0.52, 0.53$ and 0.54. The dash dot line (blue) in 
the left plot represents Brillouin curve $K_\parallel M(T)$ for spin $1/2$; the 
solid line (red) is theoretical $4\pi M(T)$ curve determined in 
\protect\cite{6} within the framework of the spin-fluctuation model \protect\cite{3}. Crosses in 
the right plot represent $K_\parallel M(T)$ dependence for $x \approx 0.55$. The 
insets show experimental temperature dependencies of the anisontropy constant $K_1$.}
\label{fig3}
\end{figure*}

Temperature dependencies of the $K_\parallel M$ parameter obtained by means of 
Eq.(\ref{fmr1}) for the films with Mn concentration $x \approx 0.52 - 0.55$ are given in 
Fig.\ref{fig3}. For all the samples, the low temperature value of the $K_\parallel M$ 
parameter is about 10~kOe (see Table~\ref{table}). The $K_\parallel M(T)$ curve for 
the sample with $x \approx 0.52$ has the Brillouin-like shape with 
$T_\text{C} \approx 300$~K. Note that the Brillouin curve gives smaller 
$T_\text{C}$ value than found from static magnetization measurements \cite{6} 
(more precisely $M(T)$ dependence can be fitted within 
spin-fluctuation model \cite{3}; see Fig.\ref{fig3} and \cite{6}). For the films 
with $x \geq 0.53$, the curve $K_\parallel M(T)$ is closer to the linear 
dependence: this fact can be caused by a heterogeneity of samples that is also 
confirmed by a larger line width of the samples with $x \approx 0.53 - 0.55$ in 
comparison with the case $x \approx 0.52$ (see the inset in Fig.\ref{fig1}).

\begin{table}[b]
\centering
\caption{Effective demagnetization, anisotropy fields and corresponding 
anisotropy constants deduced for samples with Mn content $x \approx 0.52$, 0.53 and 
0.54 at $T = 4.2$~K using Eqs.(\protect\ref{fmr1}--\protect\ref{fmr2a}).}
\vskip 3mm
\renewcommand{\arraystretch}{1.5}
\begin{tabular}{|c|ccc|cc|}
\hline
~~~$x$~~~ & $4\pi M$ & $K_\parallel M$ & $K_\perp M$ & ~~~~$K_1$~~~~ & $K_2$\\
& \multicolumn{3}{c|}{(kOe)} & \multicolumn{2}{c|}{($10^6$\,erg/cm$^3$)}\\ \hline
0.52 & 5.5 & 9.0 & 6.4 & $0.8$ & $-0.3$\\
0.53 & 4.1 & 10.8 & 8.7 & $1.1$ & $-0.2$\\
0.54 & 3.9 & 9.5 & no data & $0.9$ & no data\\
\hline
\end{tabular}
\label{table}
\end{table}

Besides the experimental $K_\parallel M(T)$ curves, Fig.\ref{fig3} represents the 
temperature dependencies of the demagnetizing field $4\pi M$ calculated from static 
magnetization data at $H = 10$~kOe applied in the film plane. The shapes of the 
$K_\parallel M(T)$ and $4\pi M(T)$ curves are close to each other, while the low 
temperature values of $K_\parallel M$ parameter exceed $4\pi M$ considerably 
(about two times; see Table~\ref{table}). These two values equalize only in the 
vicinity of $T_\text{C}$. In frame of the phenomenological model 
exposed in \ref{appA}, the large difference between $K_\parallel M$ and $4\pi M$ 
can be explained by the fact that the crystal 
structure driven second order easy plane magnetic anisotropy is comparable with the 
sample shape magnetic anisotropy.  The similarity of the $M(T)$ and $K_\parallel M(T)$ 
curves means that $K_\parallel$ is almost temperature independent and consequently 
$K_1(T) \sim M(T)^2$. The inset in Fig.\ref{fig3} demonstrates the resulting temperature 
dependence of the $K_1$ constant.

\begin{figure}[h]
\centering
\includegraphics[width=0.8\columnwidth]{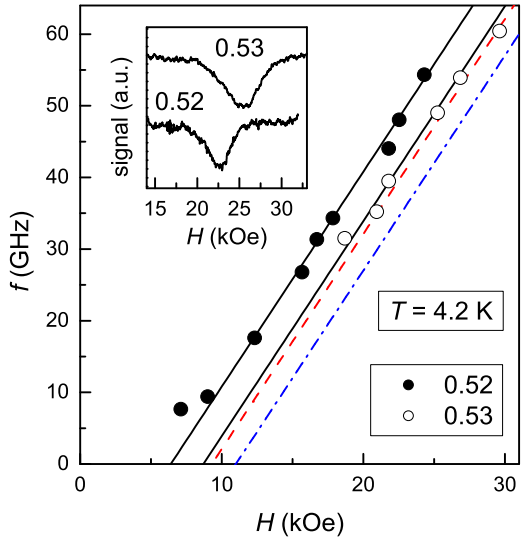}
\caption{Dependence of the resonance frequency on the magnetic field 
applied normal to the film plane for samples with $x = 0.52$ and $x = 0.53$ at $T = 
4.2$~K. Points are experimental data; solid lines are the theoretical curves 
according to Eq.(\protect\ref{fmr2}). The dashed (red) and dash-dotted (blue) lines would 
correspond to the samples with $x = 0.52$ and $x = 0.53$ respectively if we considered 
$K_\perp M = K_\parallel M$ (see details in text). The inset demonstrates examples of the 
experimental resonance spectra at frequency 49~GHz.}
\label{fig4}
\end{figure}

To obtain further insight into the peculiarities of the magnetic anisotropy of 
our system, the $f(H)$ dependencies were investigated for samples with $x \approx 
0.52$ and $x \approx 0.53$ at $T = 4.2$~K, when the applied field was 
perpendicular  to the film plane (Fig.\ref{fig4}).

Within the phenomenological model exposed in \ref{appA}, in this case the FMR 
frequency $f_\perp$ in the saturation regime has a linear dependence on the applied 
external field:
\begin{equation}
f_\perp=\gamma(H-K_\perp M),
\label{fmr2}
\end{equation}
where $K_\perp$ is the effective hard axis magnetic anisotropy coefficient, 
$K_\perp M$ is the effective field of hard axis magnetic anisotropy. 
Due to this anisotropy, the FMR line in perpendicular geometry is shifted to 
the region of larger fields comparing the paramagnetic case $f=\gamma H$. This behavior 
is opposite to the case of the parallel geometry (Eq.(\ref{fmr1})) where the FMR line was 
shifted to lower fields (Fig.\ref{fig1}). Eq.(\ref{fmr2}) is applicable in magnetic fields exceeding 
the saturation field $H_\text{S}=K_\perp M$ while below this value $f_\perp=0$.

Without the crystal structure driven magnetic 
anisotropy contribution, $K_\perp=K_\parallel=4\pi$ and Eq.(\ref{fmr2}) transforms into 
the Kittel formula for the FMR frequency, when applied field $H$ lies normally 
to the thin film plane \cite{13}.
If we take into account only the second order crystal 
structure driven magnetic anisotropy contributions to describe the total 
magnetic anisotropy in our system, the coefficient $K_\perp$ coincides with 
$K_\parallel$ \cite{12}.

In agreement with Eq.(\ref{fmr2}), the experimental $f_\perp(H)$ dependencies are linear in the 
region of high frequencies and fields (Fig.\ref{fig4}). Nevertheless, the $K_\perp M$ 
parameter differs from the $K_\parallel M$ parameter. It is seen from Fig.\ref{fig4} 
that the experimental $f_\perp(H)$ dependencies are poorly described using for the
$K_\perp M$ parameter the $K_\parallel M$ value obtained in the parallel geometry.
For both samples, the 
$K_\perp M$ parameter is less than $K_\parallel M$ but exceeds the $4\pi M$ 
value obtained from static magnetization measurements (see Table~\ref{table}). 

Following the simplest phenomenological approach of \linebreak\ref{appA}, the difference between 
$K_\perp$ and $K_\parallel$ can be attributed to the effect of a higher order 
uniaxial magnetic anisotropy of the sample. Here the term "uniaxial anisotropy" 
means that the magnetic energy expression has a uniaxial symmetry, i.e. it is 
invariant with respect to arbitrary rotations about the axis $z$ normal to the sample 
plane. The anisotropy order is defined by the power of $z$-component of magnetization 
in the energy expression.

Taking into account the second order 
easy plane anisotropy and neglecting higher than fourth order terms of uniaxial 
anisotropy (see \ref{appA}), the relation between the $K_\perp M$ and 
$K_\parallel M$ effective fields has the form:
\begin{equation}
K_\perp M=K_\parallel M + \frac{4K_2}{M},
\label{fmr2a}
\end{equation}
where $K_2$ is the fourth-order constant of crystal structure driven magnetic 
anisotropy. Thus, Eqs.(\ref{fmr1}--\ref{fmr2a}) are similar to those used in Ref.~\cite{14} with 
a little different definition of the $K_1$ and $K_2$ constants.

Note that a more complex description by means of two independent fourth-order 
phenomenological constants is also possible by presuming a tetragonal-like 
character of the film distortion. However, it is beyond the scope of the current 
paper.

The obtained low-temperature values of the $K_1$ and $K_2$ constants are given 
in Table~\ref{table}. The positive sign of the $K_1$ constant corresponds to a 
second-order easy plane contribution into the total magnetic anisotropy of the 
film. The negative sign of the $K_2$ constant corresponds to a fourth-order easy 
axis contribution into the total magnetic anisotropy of the film.

\begin{figure}
\centering
\includegraphics[width=0.9\columnwidth]{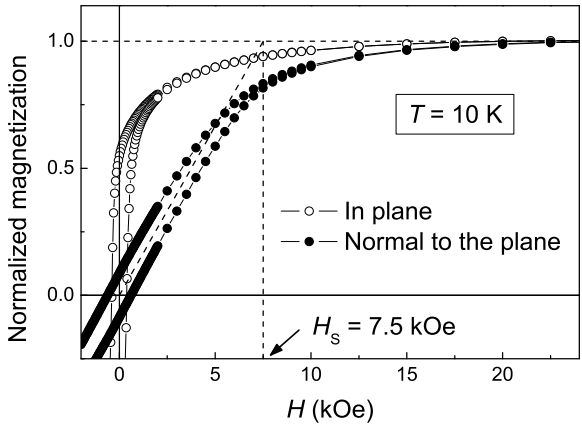}
\caption{Hysteresis loops of the sample with $x = 0.52$ in case of the field 
applied in the film plane and normal to it at $T = 10$~K.}
\label{fig5}
\end{figure}

As an additional demonstration of the role of magnetic anisotropy effects in our 
system, the low-temperature magnetization curve for the sample with $x \approx 
0.52$ was measured in the field applied perpendicular to the sample plane 
(Fig.\ref{fig5}). If only the second-order easy plane anisotropy takes place, the 
magnetization must depend on the magnetic field linearly below the saturation 
field $H_\text{S} = K_\parallel M$. The anisotropy of higher orders results in a 
nonlinearity of the magnetization curve. In this case the saturation field is 
defined by $H_\text{S} = K_\perp M$.

The experimental magnetization curve (Fig.\ref{fig5}) shows a \linebreak
smooth approach to 
saturation. Moreover, there is noticeable hysteresis of the $M(H)$ curve (the 
coercive field $H_\text{C} \approx 0.65$~kOe). Linear extrapolation of the 
initial part of the magnetization curve leads to saturation field $H_\text{S} 
\approx 7.5$~kOe. The obtained value is smaller than the $K_\parallel M$ 
observed in FMR experiments, but larger than the $K_\perp M$. Thus, there is 
only some qualitative agreement of static magnetization and resonance data. The 
static magnetization curves confirm the existence of significant second-order 
easy plane anisotropy, but allow for neither confirmation nor rejection our 
assumption about the effect of the higher order contributions of crystal 
structure driven magnetic anisotropy on the total magnetic anisotropy of the 
system.

\subsection{Structure investigations}

\begin{figure}
\includegraphics[width=\columnwidth]{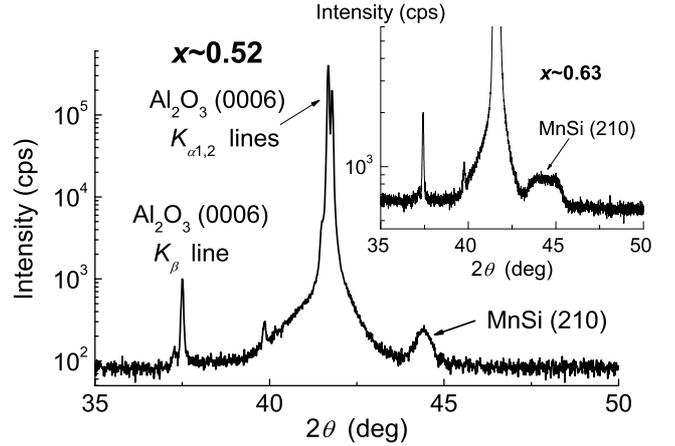}
\centering
\caption{The X-ray diffraction pattern for the Mn$_x$Si$_{1-x}$/Al$_2$O$_3$ 
structure with $x \approx 0.52$. The inset shows the XRD curve for sample with 
$x \approx 0.63$.}
\label{fig6}
\end{figure}

The results of X-ray diffraction measurements for the\linebreak
Mn$_x$Si$_{1-x}$/Al$_2$O$_3$(0001) structure ($x \approx 0.52$) are shown in 
Fig.\ref{fig6}. The diffraction curve contains strong reflection peaks from\linebreak
Al$_2$O$_3$(0006): $2\theta = 41.68^\circ$ for the CuK$_{\alpha1}$ line, 
$2\theta = 41.78^\circ$ for the CuK$_{\alpha2}$ line, and $2\theta = 37.5^\circ$ 
for the line CuK$_{\beta1}$. In addition to these peaks, this curve contains a broad 
peak from the $\varepsilon$-MnSi(210) film with B20 structure for the CuK$_\alpha$ 
line at $2\theta = 44.43^\circ$.

The integral characteristic of the film's structural quality is the rocking curve 
width at half maximum (FWHM$_\omega$). For the film under study, the FWHM$_\omega$ parameter 
at $2\theta = 44.43^\circ$ is $\Delta\omega \approx 0.4^\circ$, whereas the 
FWHM$_\omega$ value for the single-crystalline film of the same thickness should be 
about 250 seconds of arc. Such a broad peak is a signature of a pronounced mosaicity 
and imperfection of the film structure, in particular, caused by the lattice constant 
mismatch between Al$_2$O$_3$ substrate and MnSi film as well as by the Mn excess (for 
MnSi with B20 structure, the lattice constant $a \approx 4.56$~\AA; for Al$_2$O$_3$ 
$a \approx 4.76$~\AA). At increasing Mn content the $\varepsilon$-MnSi(210) peak 
transforms to a "flat hill" about $2^\circ$ wide at $x \approx 0.63$ (see inset 
in Fig.\ref{fig6}).

\begin{figure}
\includegraphics[width=0.8\columnwidth]{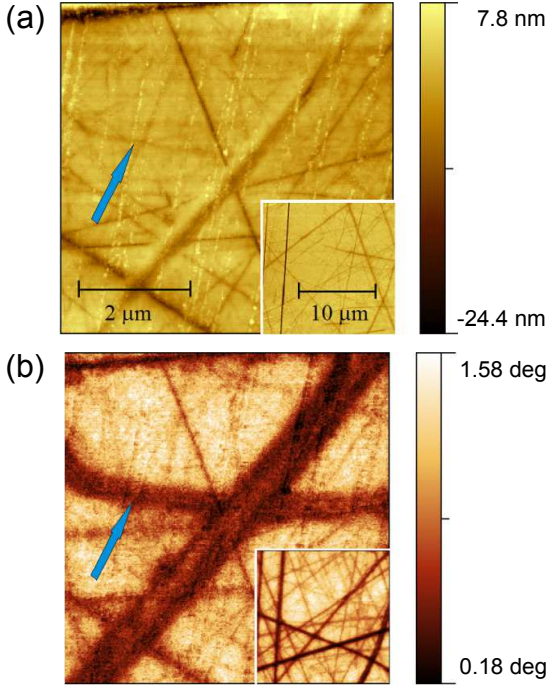}
\centering
\caption{AFM (a) and MFM (b) images for the Mn$_x$Si$_{1-x}$ 
film with $x \approx 0.52$ and $T_\text{C} \approx 330$~K. The light regions 
in the MFM image are the attracting magnetic areas. Dark regions are areas in which 
there is no MFM probe attraction. The arrow shows the interface between 
crystallites in case when it is poorly appeared in AFM mode (a) 
while it is obvious in MFM mode (b). Insets in Figs.\protect\ref{fig7}a and b demonstrate the fragments 
of AFM and MFM images, respectively, at large film surface scanning area 
$20\times20~\mu\text{m}^2$.}
\label{fig7}
\end{figure}

To obtain further insight into peculiarities of the film structure, we analyze 
AFM and MFM images of the sample surface (see Fig.\ref{fig7}). The AFM and MFM measurements were 
performed in ambient conditions for the Mn$_x$Si$_{1-x}$ film with $x \approx 0.52$ 
and $T_\text{C} \approx 330$~K (magnetization data are shown in Fig.\ref{fig3}). For 
receiving the MFM images (Fig.\ref{fig7}b), the two-pass technique (lift-mode) was used. 
The height of lift on the second pass was about $30-50$~nm. Change of the probe 
oscillation phase was recorded at the fixed pump frequency. The light regions on the 
MFM images correspond to an increase of the phase arising at a reduction of the 
probe resonant frequency which is caused by its attraction to the surface. 
Therefore, light regions on the MFM images display the areas where the probe is 
attracted to the sample, and dark strips show the areas where such the 
attraction becomes weak or is absent. Magnetic images do not depend on the 
sample previous history, i.e. they do not change at preliminary magnetization of 
the sample in this or that direction. The saturation magnetization of the sample 
under conditions of measurements ($T \approx 290$~K) is about 100~emu/cm$^3$, 
and its coercitivity is a fortiori less than 50~Oe (see Fig.\ref{fig3} and 
Ref.~\cite{6}). In this situation, the local reversal magnetization of the 
sample in the field of MFM probe is possible, leading to its attraction.

According to AFM results, 
the depth of weakly pronounced inter-block interfaces (thin lines in Fig.\ref{fig7}a) does not 
exceed 2~nm, while strongly pronounced inter-crystallite interfaces (thick 
lines) have the form of cavities with the depth of \linebreak$<10$~nm at the 70~nm 
thickness of the film. Comparison of AFM and MFM images shows that the positions 
of inter-crystal\-lite interfaces correlate on the whole with dark strips in MFM 
images (Fig.\ref{fig7}b). However, the width of these strips ($\sim0.5~\mu$m) 
considerably exceeds the width of lines ($\leq 0.1~\mu$m) separating blocks 
(Fig.\ref{fig7}a) in the AFM images. Moreover, there exist inter-crystallite 
interfaces which are almost not revealed in AFM images (shown by arrow in 
Fig.\ref{fig7}a) in which, however, magnetic heterogeneity (dark strips in Fig.\ref{fig7}b) is 
brightly expressed.

\subsection{Possible origin of magnetic anisotropy}

The itinerant cubic ferromagnet singe-crystal MnSi with B20 structure has a 
weak forth-order cubic magnetic anisotropy. But in case of a thin epitaxial 
MnSi film deposited on a thick substrate, the induced uniaxial magnetic anisotropy 
can be essential due to the strain caused by lattice mismatch between the film 
and the substrate (see \cite{15} and references therein). 

In our case, sufficiently large mismatch ($\approx4$\,\%) between the Al$_2$O$_3$ 
substrate and Mn$_x$Si$_{1-x}$ film is realized. It is one of the main reasons 
for polycrystallinity of the grown film and thus initiates an existence of 
inter-crystallite and crystallite-substrate strain, producing crystal twin planes 
or inter-crystallite boundaries \cite{16} (below we will use a unified term 
"plane defect"). 

XRD measurements clearly show that Mn$_x$Si$_{1-x}$ crystallites are ordered 
normally to the surface of Al$_2$O$_3$ substrate, i.e. the grown films are textured 
(have a mosaic type). According to AFM-MFM images, the characteristic size of 
crystallite is about 1~$\mu$m. Obviously, in the frame of used methods we are unable 
to adduce direct experimental proofs of the strain inside our films; so, our supposal 
should be verified in future studies. 
However, it is well known that mechanical strain on the plane 
defect can induce elastic or plastic deformation (even dislocation) near 
this defect \cite{16}. Following this paradigm, at least a part of thin lines 
(one of them is indicated in Fig.\ref{fig7}a by arrow) may be possibly associated with 
the projections of plane defects with elastic strains on the 
film surface. Some lines (not shown in Fig.\ref{fig7}) have a profound cavity shape and 
may be possibly attributed to the plane defects with strong non-elastic 
deformations or dislocations in the film.

The MFM data shed light on the magnetic structure of the film surface, while their 
interpretation seems to be ambiguous. A single-domain thin film with strong 
easy-plane anisotropy does not exhibit local reversal magnetization in the field 
of MFM probe, so the MFM foreground color should look like homogeneously dark. 
However, in our experiments the MFM signal is locally dark only nearby the lines 
possibly corresponding to projections of plane defects, as the probe approaches 
them, while it remains light far from them. A possible reason for such the 
effect is due to the significant enhancement of magnetic anisotropy near the 
plane defects in the film. This enhancement may be provided by an increase of 
anisotropic (for example, spin-orbit) component of effective exchange coupling 
between local magnetic moments of Mn-containing nanometer scale defects, due to 
strong crystal potential distortions near the plane defect in the 
nonstoichiometric Mn$_x$Si$_{1-x}$ alloy. 
Following our supposition, the "local" axes of magnetic anisotropy are oriented 
normally to the plane defect, i.e. lie in most part in the film plane. The plane 
defects are randomly distributed in the film, they strongly pin local magnetic 
moments of Mn-containing defects and block a local reversal magnetization in MFM 
measurements.

The plane defect driven magnetic anisotropy mechanism does not qualitatively 
contradict to the above performed FMR data and their interpretation in frame of 
the phenomenological description of \ref{appA}. In \ref{appB}, we propose a 
simple quantum mechanical model of randomly distributed plane defects having 
spin-orbit coupling with the matrix of a weak itinerant ferromagnet. By means of 
this model, we support the phenomenological approach of \ref{appA}.
In particular, the model predicts essential second order easy plane anisotropy 
contribution with $K_1 > 0$ and the additional forth order uniaxial anisotropy 
contribution with $K_2 < 0$. However, the ratio $|K_2|/K_1$ estimated from the 
model is $|K_2|/K_1 \sim 10^{-3}-10^{-4}$; this is much less than the value 
$|K_2|/K_1 \sim 0.2-0.4$ found from the experiment (see Table~\ref{table}). The possible 
reason of this disagreement is the used perturbation approach (see \ref{appB}) as 
well as the strain between crystallites and the substrate which is neglected in the 
proposed model. At the same time, it should be kept in mind that phenomenological 
description of \ref{appA} is developed for a purely homogeneous case, i.e. does not 
consider distribution of local anisotropy in the plane and on the film thickness. 
Therefore, the constants $K_1$ and $K_2$ found with its help have only an efficient 
character.

\section{Conclusions}

In this work, for thin films of nonstoichiometric Mn$_x$Si$_{1-x}$ alloys with 
different manganese content ($x \approx 0.44 - 0.63$) the FMR measurements were 
performed in the wide range of frequencies ($f = 7 - 60$~GHz) and magnetic 
fields ($H = 0 - 30$~kOe). For samples with $x \approx 0.52 - 0.55$, the FMR 
data confirm the early reported FM order with high Curie temperatures 
$T_\text{C} \sim 300$~K \cite{6,7}. Earlier, we have explained the appearance of 
such FM order in Mn$_x$Si$_{1-x}$ films in frame of a non-conventional 
defect-induced carrier-mediated mechanism~\cite{3}.

Further to the fact of FM order itself, studied samples also demonstrated in FMR 
measurements an intricate character of magnetic anisotropy, which can be 
described in a phenomenological way as a combination of two contributions: the 
second order easy plane anisotropy component and the forth order uniaxial 
anisotropy component with easy direction normal to the film plane. 
In frame of above mentioned assumption we attribute this magnetic anisotropy 
to the existence of a well-pronounced mosaic (polycrystalline) structure of the films. 
We believe that such a mosaic structure is revealed in presented XRD and AFM 
measurements, be accompanied by the strain between crystallites and/or 
crystallites-substrate. Following our model, these local strain can initiate an 
enhancement of the spin-orbital anisotropic component of exchange interaction between 
the local moment on the point magnetic defect and itinerant electron spin in the 
matrix (see \ref{appB}). This enhancement becomes apparent as a pinning of local 
magnetic moments in the MFM images.

We hope that the combination of FMR, XRD, AFM and MFM methods showed its efficiency 
in the study of nonstoichiometric Mn$_x$Si$_{1-x}$ alloys as a new class of 
high-temperature FM semiconductors with unusual properties.

\section*{Acknowledgments}

Authors are grateful to Dr.~Vasiliy Glazkov for assistance in carrying out 
low-temperature FMR measurements in the region of high fields. We express our 
gratitude to Dr. Alexei Temiryazev for AFM and MFM measurements. We would also 
like to thank him as well as Dr.~Dmitry Kholin and Dr. Elkhan Pashaev for 
fruitful discussions of FMR and XRD results.

The work was partly supported by the RFBR (grant Nos. 15-07-01170, 14-07-91332, 
14-07-00688, 13-07-00477, 15-29-01171, 14-47-03605, 15-07-04142), NBICS Center 
of the Kurchatov Institute. The work at HZDR is financially supported by DFG (ZH 
225/6-1)

\appendix

\section{FMR in magnetic film with perpendicular anisotropy}\label{appA}

Taking into account the perpendicular uniaxial anisotropy, the magnetic energy 
density of a ferromagnetic film is given by the following expression:
\begin{equation}
E = -\mathbf{H\cdot M} + 2\pi M_z^2 + E_A(M_z),
\end{equation}
where the first term is Zeeman energy in magnetic field, the second term is 
shape anisotropy of the sample ("demagnetization energy") and the last term 
represents general form of uniaxial anisotropy energy with the anisotropy axis 
oriented along vector \textbf{z} normal to the film plane.

Neglecting dissipation, the precession of the magnetic moment is determined by 
the Landau-Lifshitz equation:
\begin{equation}
\frac{\partial\mathbf{M}}{\partial t} = \gamma [\mathbf{M} \times 
\mathbf{H}_\text{eff}],
\label{a2}
\end{equation}
where the effective field:
\begin{equation}
\mathbf{H}_\text{eff} = -\frac{\partial E}{\partial \mathbf{M}} = \mathbf{H} - 
4\pi M_z\mathbf{z} - \frac{\partial E_A}{\partial M_z}\mathbf{z}.
\end{equation}
Resonance frequency is defined as eigenfrequency of the system (\ref{a2}) after its 
linearization near equilibrium orientation of the \textbf{M} vector. In 
sufficiently large fields exceeding the saturation field, the static 
magnetization is oriented along the magnetic field 
\textbf{M}$\parallel$\textbf{H}. Taking into account this condition, in case of 
the field applied in the film plane, the resonance frequency is defined by 
Eq.(\ref{fmr1}), where:
\begin{equation}
K_\parallel M = 4\pi M + M \left(\frac{\partial^2 E_A}{\partial 
M_z^2}\right)_{M_z = 0}.
\label{a4}
\end{equation}
When the field is applied perpendicular to the film plane, the resonance 
frequency is defined by Eq.(\ref{fmr2}), where:
\begin{equation}
K_\perp M = 4\pi M + \left(\frac{\partial E_A}{\partial M_z}\right)_{M_z = M}.
\label{a5}
\end{equation}
Thus, in the presence of the uniaxial anisotropy the $K_\parallel$ and $K_\perp$ 
parameters generally speaking do not coincide. Writing the $E_A(M_z)$ function 
in the form of decomposition:
\begin{equation}
E_A = K_1 \cos^2 \theta + K_2 \cos^4 \theta + K_3 \cos^6 \theta + \cdots,
\label{a6}
\end{equation}
where $\cos \theta = M_z/M$, Eq.(\ref{a4}) transforms into (\ref{fmr1a}) and 
Eq.(\ref{a5}) takes form:
\begin{equation}
K_\perp M = 4 \pi M + \frac{2K_1}{M} + \frac{4K_2}{M} + \frac{6K_3}{M} +\cdots
\label{a7}
\end{equation}
If the sixth and higher order anisotropy constants are negligible, Eq.(\ref{a7}) 
and Eq.(\ref{fmr1a}) lead to the expression (\ref{fmr2a}).

\section{Microscopic approach}\label{appB}

Self-consistent theory of spin fluctuations in the homogeneous itinerant FM 
\cite{3} is based on the simple model Hamiltonian of interacting fermions:
\begin{align}
\label{b1}
\mathscr{H}_0 &= \sum_\alpha \int \Psi_\alpha^+(\mathbf{r}) 
\varepsilon(\mathbf{k})\Psi_\alpha(\mathbf{r}) d\mathbf{r} + \\
&+ \sum_{\alpha,\beta} \int \Psi_\alpha^+(\mathbf{r}) \Psi_\alpha(\mathbf{r}) 
U(\mathbf{r}-\mathbf{r}^\prime) \Psi_\beta^+(\mathbf{r}^\prime) 
\Psi_\beta(\mathbf{r}^\prime) d\mathbf{r} d\mathbf{r}^\prime, \nonumber
\end{align}
where $\Psi_\alpha^+(\mathbf{r})$ and $\Psi_\alpha(\mathbf{r})$ are creation and 
annihilation operators of fermions, $\varepsilon(\mathbf{k})$ is fermions 
spectrum, $\mathbf{k}=-i\partial/\partial \mathbf{r}$ is operator of quasi 
momentum, $(\mathbf{r},\mathbf{r}^\prime)$ are three-dimensional space vectors, 
$(\alpha,\beta)$ are spin indices, $U(\mathbf{r}-\mathbf{r}^\prime)$ is 
effective potential of fermions interaction. Hamiltonian (\ref{b1}) is spin-rotation 
invariant and does not contain relativistic contributions in the fermions 
spectrum and effective interaction.

Let us modify Hamiltonian (\ref{b1}) for a case of mosaic film of itinerant FM deposed 
on the non-magnetic substrate. We suppose that this film is composed of 
macroscopic grains with characteristic size $l$ in the film plane and nearest 
neighbored grains are separated one from other by a narrow interfacial region 
with characteristic thickness $d \ll l$. Notice that $(d,l)$ significantly 
exceed the lattice parameter $a$. The grain boundaries are orthogonal to the 
film plane throughout all the film thickness and form an arrow of 
two-dimensional defects inside the film. The crystal potential in the 
interfacial region significantly differs from the potential inside the grains 
due to broken chemical bonds, inter grain clustered aggregates, mechanical 
strains etc. and thus strongly modifies the fermions motion. To model the effect 
of interfacial boundaries on the fermions behavior in itinerant FM, we treat 
these boundaries as non-magnetic macroscopic plane defects embedded into the 
homogeneous matrix and introduce additional term in the Hamiltonian of our 
system:
\begin{equation}
\mathscr{H}_1 = \sum_{\alpha,\beta,,n} \int \Psi_\alpha^+(\mathbf{r}) 
\lbrace[V_n I + \mathbf{D}_n \cdot \boldsymbol{\sigma}] \delta 
(\boldsymbol{\rho}-\boldsymbol{\rho}_n)\rbrace_{\alpha\beta} 
\Psi_\beta(\mathbf{r}) d\mathbf{r}.
\end{equation}
Here $\delta(x)$ is delta-function, $\mathbf{r} = 
(\boldsymbol{\rho},\mathbf{z})$, two-dimensional vector $\boldsymbol{\rho}_n$ 
defines the $n$-th plane defect position in the film, axis \textbf{z} is 
orthogonal to the film plane, $\boldsymbol{\sigma}$ is vector composed from the 
Pauli matrices, $V_n$ and $\mathbf{D}_n$ are respectively scalar (Coulomb) and 
vector (spin-orbit) components of the  $n$-th plane defect potential. The 
component $\mathbf{D}_n$ may be expressed in the Bychkov-Rashba form as 
$\mathbf{D}_n = \eta[\mathbf{e}_n \times \mathbf{k}]$, where parameter $\eta$ is 
proportional to the fine structure constant and gradient of interface potential, 
$\mathbf{e}_n$ is unit vector normal to the $n$-th defect plane \cite{3}.

Effect of the bulk magnetic defects with local spins 
$\lbrace\mathbf{S}_j\rbrace$ on the behavior of fermions in itinerant FM is 
traditionally considered within the Hamiltonian:
\begin{equation}
\mathscr{H}_2 = \sum_{\alpha,\beta,,j} \int \Psi_\alpha^+(\mathbf{r}) \lbrace 
[J_j \mathbf{S}_j \cdot 
\boldsymbol{\sigma}]\delta(\mathbf{r}-\mathbf{r}_j)\rbrace_{\alpha\beta} 
\Psi_\beta(\mathbf{r}) d\mathbf{r}.
\end{equation}
Here $J_j$ is corresponding exchange coupling integral, local spins 
$\lbrace\mathbf{S}_j\rbrace$ are randomly distributed in the three-dimensional 
fermions matrix with the mean inter-spin distance $b$, $a \ll b \ll (d,l)$.

From the total Hamiltonian of the system, ~$\mathscr{H} = \mathscr{H}_0 + 
\mathscr{H}_1 + \mathscr{H}_2$, it is possible to obtain the free energy 
functional $\Phi\lbrace \mathbf{m}(\mathbf{r}), \mathbf{S}(\mathbf{r})\rbrace$ 
of itinerant FM with both 2D macroscopic non-magnetic defects and microscopic 3D 
magnetic defects as a series expansion in terms of the magnetic moment densities 
of itinerant fermions $\mathbf{m}(\mathbf{r})$ and local spins 
$\mathbf{S}(\mathbf{r})$. The form of $\Phi\lbrace \mathbf{m}(\mathbf{r}), 
\mathbf{S}(\mathbf{r})\rbrace$ depends on the studied temperature region and 
relative contributions of different terms in the Hamiltonian $\mathscr{H}$. 
Early in Refs.~\cite{3}, we analyzed $\Phi\lbrace \mathbf{m}(\mathbf{r}), 
\mathbf{S}(\mathbf{r})\rbrace$ at the high-tem\-pe\-ra\-ture region $T > T_\text{C}$, 
where $T_\text{C}$ is the global Curie temperature of the system, completely 
neglecting the term $\mathscr{H}_2$, i.e. preserving spin-rotation invariance. 
Moreover, we revealed in Refs.~\cite{3} a drastic enhancement of exchange 
coupling between local spins of defects due to the effect of itinerant fermions 
spin fluctuations and derived corresponding expression for $T_\text{C}$.

In this paper, we analyzed $\Phi\lbrace \mathbf{m}(\mathbf{r}), 
\mathbf{S}(\mathbf{r})\rbrace$ below the global Curie temperature, in the 
temperature range $T_\text{SF} < T < T_\text{C}$, where $T_\text{SF}$ is 
characteristic temperature of the freezing of itinerant fermions spin 
fluctuations. Without local spins, i.e. in the purely itinerant FM, 
$T_\text{SF}$ should be the Curie temperature \cite{3}. In the temperature 
region under consideration, the mean field approximation for the spin densities 
$\mathbf{m}(\mathbf{r}) = \mathbf{m}$ and $\mathbf{S}(\mathbf{r}) = \mathbf{M}$ 
seems to be reasonable on the spatial scale exceeding the characteristic lengths 
$(b,d,l)$ of both micro- and macroscopic defects. Thus, taking into account all 
the terms in the Hamiltonian $\mathscr{H}$, after the micro- and macroscopic 
defect distribution averaging, we can obtain an expression for $\Phi\lbrace 
\mathbf{m}, \mathbf{M}\rbrace$ as a series expansion in terms of \textbf{m} and 
\textbf{M}. Generally speaking, this expression is cumbersome, but here we take 
interest only to the terms in $\Phi\lbrace \mathbf{m}, \mathbf{M}\rbrace$ 
dependent on the directions of vectors \textbf{m} and \textbf{M}. Let us assume 
that averaged local spin density \textbf{M} is saturated in the considered 
temperature region and does not change its absolute value $M$. In that case we 
can simplify our analysis omitting in $\Phi\lbrace \mathbf{m}, 
\mathbf{M}\rbrace$ the terms independent of the \textbf{M} direction. In the 
fourth order in \textbf{m} and first order in \textbf{M}, we can show using 
conventional diagram techniques for the free energy functional \cite{3}, that:
\begin{align}
\label{b4}
\Delta\Phi\lbrace \mathbf{m}, \mathbf{M}\rbrace &= \Phi\lbrace \mathbf{m}, 
\mathbf{M}\rbrace - \Phi_0\lbrace\mathbf{M}\rbrace =  \nonumber \\
&= \Phi_0\lbrace\mathbf{m}\rbrace + \Phi_1\lbrace\mathbf{m}\rbrace + \delta 
\Phi\lbrace \mathbf{m}, \mathbf{M}\rbrace, \\
\Phi_0\lbrace\mathbf{m}\rbrace \approx A_1 & m^2 + A_2 m^4,~~ 
\Phi_1\lbrace\mathbf{m}\rbrace \approx B_1 m_z^2 + B_2 m_z^4, \nonumber \\
&\delta \Phi\lbrace \mathbf{m}, \mathbf{M}\rbrace \approx \lambda \mathbf{m} 
\cdot \mathbf{M}. \nonumber
\end{align}
The $\Phi_0\lbrace\mathbf{M}\rbrace$ term simply shift the free energy scale and 
does not takes interest for us, $\Phi_0\lbrace\mathbf{m}\rbrace$  and  
$\Phi_1\lbrace\mathbf{m}\rbrace$ are respectively isotropic and anisotropic in 
\textbf{m} contributions of itinerant fermions, $\delta \Phi\lbrace \mathbf{m}, 
\mathbf{M}\rbrace$ is contribution of exchange coupling between itinerant 
fermions spins and magnetic defects local spins. The coefficients in formulas 
(\ref{b4}) can be estimated in the temperature region $T_\text{SF} < T < T_\text{C}$ 
(see Refs.~\cite{3}) as:
\begin{equation}
\begin{split}
&A_1 \approx W^{-1}(\nu_\text{F} Q_\text{SF}/W)^2 [\zeta(0)/\zeta(T)]^2, \\
&A_2 \approx W^{-3}, \qquad \lambda \approx J (a/b)^3 W^{-2},
\end{split}
\label{b5}
\end{equation}
\begin{equation}
\begin{split}
&B_1 \approx W^{-1}(d/l)(\eta/\nu_\text{F})^2, \\
&B_2 \approx -W^{-3}(d/l)(\eta/\nu_\text{F})^4.
\end{split}
\label{b6}
\end{equation}
Here $W$ is energy scale of the order of fermions bandwidth, $\nu_\text{F}$ is 
the Fermi velocity, $Q_\text{SF}$ is the cut-off wave vector of itinerant 
fermions spin fluctuations, $\zeta(0) \approx \nu_\text{F}/W$, $\zeta(T)$ is 
correlation length of itinerant fermions spin fluctuations, renormalized by a 
scattering on the Coulomb component of macroscopic defects potential $V$. This 
scattering is not explicitly included in formulas (\ref{b5}), since at considered 
temperatures and relations between the parameters of our model $(d/l) \ll 1$, 
$[d/\zeta(0)] \ll 1$, $(V/W)^2 \ll 1$ it does not lead to new physical effects.

It is seen from Eq.(\ref{b5}) and Eq.(\ref{b6}) that coefficients $A_1$, $A_2$, $B_1$, 
$\lambda$ are positive in the considered temperature region, while coefficient 
$B_2$ is negative. Formally this fact is due to an interplay\- between different 
diagrams of the eighth order in $\Phi\lbrace \mathbf{m}, \mathbf{M}\rbrace$~
series expansion in the effective perturbation field 
$\left[\mathbf{m}(\mathbf{r})+\sum_n 
\mathbf{D}_n(\boldsymbol{\rho}-\boldsymbol{\rho}_n)\right]$, leading to 
appearance of the fourth order in \textbf{m} anisotropic contributions to 
$\Phi_1\lbrace\mathbf{m}\rbrace$ after averaging over  macroscopic defects 
distribution. We do not attribute a profound physical meaning to this result, 
while it seems noteworthy.

Varying $\Delta \Phi \lbrace \mathbf{m}, \mathbf{M}\rbrace$ over \textbf{m}, 
after neglecting for simplicity the $\Phi_1\lbrace\mathbf{m}\rbrace$ 
contribution, we get in the mean field approximation for an equilibrium value  
$\mathbf{m}_0 \approx -\lambda/2A_1 \mathbf{M} [1-(\lambda A_2/A_1)^2 M^2]$. 
Substituting $\mathbf{m} = \mathbf{m}_0$ in $\Delta \Phi \lbrace \mathbf{m}, 
\mathbf{M}\rbrace$ and separating isotropic and anisotropic terms, we obtain for 
the magnetic anisotropy energy $E_A$ the expression (\ref{a6}) with $\cos\theta=M_z/M$ 
and
\begin{equation}
K_1 \sim B_1 m_0^2 > 0,\qquad K_2 \sim B_2 m_0^4 < 0.
\label{b7}
\end{equation}
Estimations of relative and a fortiori absolute values of $K_1$ and $K_2$ are 
uninformative in our model due to their very rough  approximate character. 
Obviously, in our approach contribution of the forth order terms proportional to 
$K_2$ must be small compared with contribution of  the second order terms  
proportional to $K_1$ in Eqs.(\ref{b7}) due to the used perturbation theory in the 
fine structure parameter:
\begin{equation}
|K_2|/K_1 \sim (\eta/\nu_\text{F})^2 W^{-2} m_0^2 \sim 10^{-3} \div 10^{-4} \ll 
1.
\label{b8}
\end{equation}

From the phenomenological model of \ref{appA}, to satisfy obtained FMR results 
we have to put $|K_2|/K_1 \sim 10^{-1}$. Thus, correspondence between Eq.(\ref{b8}) and 
experimental results is not yet good. This disagreement 
is not surprising since the developed model is based on the perturbation approach 
to the spin-orbit component of coupling between itinerant electron spin of the matrix 
and the plane defect moment, i.e. the calculated coefficients $K_1$ and $K_2$ are 
obtained as the lowest terms in the corresponding series expansion of the coupling 
energy on the spin-orbit components. For real alloys the used perturbation approach 
may be incorrect, but unfortunately, in the theory of itinerant ferromagnetism there 
exist no adequate description to take strong spin-orbit coupling into account.


\section*{References}

\begin{thebibliography}{99}

\bibitem{1} S. Zhou and H. Schmidt, Materials 3 (2010) 5054.
\bibitem{2} S. Zhou, K. Potzger, G. Zhang, A. M\"ucklich, F. Eichhorn, N. 
Schell, R. Gr\"otzschel, B. Schmidt, W. Skorupa, M. Helm, J. Fassbender, D. 
Geiger, Phys. Rev. B 75 (2007) 085203.
\bibitem{3} V.N. Men'shov, V.V. Tugushev, S. Caprara, Phys. Rev. B 
83 (2011) 035201;
V. N. Men'shov and V.V. Tugushev, J. Exp. Theor. Phys. 
113 (2011) 121.
\bibitem{4} S. Kahwaji, R.A. Gordon, E.D. Crozier, S. Roorda, M.D. Robertson, J. 
Zhu, T.L. Monchesky, Phys. Rev. B 88 (2013) 174419.
\bibitem{5} E.S. Demidov, V.V. Podol'skii, V.P. Lesnikov, E.D. Pavlova, A.I. 
Bobrov, V.V. Karzanov, N.V. Malekhonova, A.A. Tronov, JETP Lett. 100 
(2015) 719.
\bibitem{6} V.V. Rylkov, S.N. Nikolaev, K.Yu. Chernoglazov, B.A. Aronzon, K.I. 
Maslakov, V.V. Tugushev, E.T. Kulatov, I.A. Likhachev, E.M. Pashaev, A.S. 
Semisalova, N.S. Perov, A.B. Granovskii, E.A. Gan'shina, O.A. Novodvorskii, O.D. 
Khramova, E.V. Khaidukov, V.Ya. Panchenko, JETP Letters 96 (2012) 255.
\bibitem{7} V.V. Rylkov, E.A. Gan'shina, O.A. Novodvorskii, S.N. Nikolaev, A.I. 
Novikov, E.T. Kulatov, V.V. Tugushev, A.B. Granovskii, V.Ya. Panchenko. 
Europhys. Lett. 103 (2013) 57014.
\bibitem{8}	A. Yang, K. Zhang, S. Yan, S. Kang, Y. Qin, J. Pei, L. He, H. Li, 
Y. Dai, S. Xiao, Y. Tian, J. Alloy. Compd. 623  (2015) 438.
\bibitem{9} S.M. Stishov and A.E. Petrova, Phys. Usp. 54 (2011) 1117.
\bibitem{10} J.A. Hagmann, K. Traudt, Y.Y. Zhou, X. Liu, M. Dobrowolska, J.K. 
Furdyna, J. Magn. Magn. Mater. 360 (2014) 137.
\bibitem{11} S.V. Demishev, V.V. Glushkov, I.I. Lobanova, M.A. Anisimov, V.Yu. 
Ivanov, T.V. Ishchenko, M.S. Karasev, N.A. Samarin, N.E. Sluchanko, V.M. Zimin, 
A.V. Semeno, Phys. Rev. B 85 (2012) 045131.
\bibitem{12} S.V. Vonsovskii, \textit{Ferromagnetic Resonance} (Pergamon Press, 
Oxford, 1966).
\bibitem{13} C. Kittel, J. Phys. Radium. 12 (1951) 291.
\bibitem{14} J.-M.L. Beaujour, W. Chen, K. Krycka, C.-C. Kao, J.Z. Sun, A.D. 
Kent, Eur. Phys. J. B 59 (2007) 475.
\bibitem{15} M.N. Wilson, A.B. Butenko, A.N. Bogdanov, and T.L. Monchesky, 
Phys. Rev. B 89 (2014) 094411.
\bibitem{16} A.M. Kosevich, V.S. Boiko, Sov. Phys. Usp. 14 (1971) 286.

\end{thebibliography}
\end{document}